 \documentclass[preprint2]{aastex}
\usepackage{color}

\newcommand{\cep} {CoRoT 0223989566}
\newcommand{\fu}{$f_1$}
\newcommand{\ft}{$f_3$}
\newcommand{\fdue}{$f_2$}
\newcommand{\cd} {d$^{-1}$}
\newcommand{\rdu} {$R_{21}$}
\newcommand{\fdu} {$\phi_{21}$}

\newcommand{\ep}[1]{\textcolor{black}{#1}}

\shorttitle{CoRoT 0223989566, a triple-mode Cepheid in the Galaxy}
\shortauthors{Poretti, Baglin, \& Weiss}

\begin{document}


\title{THE CoRoT DISCOVERY OF A UNIQUE TRIPLE-MODE CEPHEID IN THE GALAXY}


\author{E. Poretti\altaffilmark{1}}
\affil{INAF-Osservatorio Astronomico di Brera, Via E. Bianchi 46, 23807 Merate, Italy}
\email{ennio.poretti@brera.inaf.it}
\and
\author{A. Baglin\altaffilmark{2}}
\affil{LESIA, Universit\'e Pierre et Marie Curie, Universit\'e Denis Diderot, Observatoire de Paris, \\F-92195 Meudon Cedex, France}
\and
\author{W.W. Weiss\altaffilmark{3}}
\affil{Institute of Astronomy, University of Vienna, T\"urkenschanzstrasse 17, A-1180 Vienna, Austria}



\begin{abstract}
The exploitation of the CoRoT treasure of stars observed in the exoplanetary field
allowed the detection of a unusual triple-mode Cepheid in the Milky Way, 
CoRoT 0223989566. The two modes with
the largest amplitudes and period ratio of 0.80 are identified with the first ($P_1$=1.29~d) and second
($P_2$=1.03~d) radial overtones.
The third period, which has the smallest amplitude but able to produce combination terms with the other two,
is the longest one ($P_3$=1.89~d). The ratio of 0.68 between the first-overtone period and the third period is the
unusual feature.
Its identification with the fundamental radial or a nonradial mode is discussed with respect to similar cases
in the Magellanic Clouds. In both cases the period triplet and the respective ratios make the star unique in our Galaxy.
The distance derived from the period--luminosity relation and
the galactic coordinates put CoRoT~0223989566 in the metal-rich environment of the
``outer arm" of the Milky Way.
\end{abstract}





\section{Introduction}

Double-mode Cepheids are a powerful tool to test
the stellar models of supergiants, since the simultaneous excitation of
two pulsation modes tightly constrains the physical parameters. In the 1990,
the introduction of the new OPAL opacities \citep{opal} solved the discrepancy between
the beat and pulsation masses \citep{bump}, also reconciling the evolutionary ones
\citep{jcd}.
Nowadays, the period ratio is
used to investigate the metallic content of  stellar systems  hosting double-mode
Cepheids. \citet{kate} obtained  observational evidence of the relation between
the period ratios and the metal abundances by means of high-resolution
spectroscopy of galactic double-mode Cepheids. The metallicity of the Large (LMC) and
Small (SMC) Magellanic Clouds has been investigated by using the period ratios of
hundreds of double-mode Cepheids in the framework of large-scale surveys:
MACHO \citep[e.g., ][]{alcock}, EROS-2 \citep[e.g., ][]{eros}, OGLE-III
\citep[e.g., ][]{soszinski2008a,soszinski2008b}

The involved radial modes are the fundamental ($F$) and  the first
($1O$), second ($2O$), and third ($3O$) overtones. Typical period ratios
are $1O/F$=0.71, $2O/1O$=0.80, $3O/1O$=0.68.
However, the large-scale surveys
 revealed other particular subclasses \citep[for a review see ][]
{pavel}: stars with a nonradial mode in close
proximity to the dominant $1O$ mode and  double-mode pulsators with
a period ratio of 0.60-0.64.

On the other hand, triple-mode Cepheids are extremely rare. 
The current statistics \citep{pavel} 
report six cases of $1O/2O/3O$ stars (three in the LMC, one in the SMC, and two in the galactic
bulge) and four cases of  $F/1O/2O$ stars (two in the LMC, two in the SMC). The two triple-mode
Cepheids in the galactic bulge have extremely short periods: the $1O$ periods are 0.295~d
and 0.230~d, corresponding to $F$ periods shorter than 0.4~d. 
\ep {The modeling of three periods and the evolutionary calculations are expected to 
place strong constraints both on Cepheid mass--luminosity
relations and on the internal physics.}

The variability of GSC~0746-01186 was discovered in the ASAS survey
\citep[ASAS~064135+0756.6; ][]{asas}. The star was classified as a Cepheid with $P$=1.28859~d,
with an amplitude of 0.41~mag and $<V>$=12.48.
Later, \citet{khruslov} identified it as a double-mode Cepheid with
periods $P_1$=1.28861~d and $P_2$=1.03153~d. These periods were confirmed by
the analysis of the NSVS data \citep{nsvs}. The ratio of 0.8005 suggested a
pulsation in the $1O$ and $2O$ modes.  
GSC~0746-01186 $\equiv$ ASAS~064135+0756.6 $\equiv$ SRa01b.16229 was
re-observed during the BEST~II survey
\citep[Berliner Exoplanet Search Telescope; ][]{best}. The double-mode
pulsation and the period ratio were both confirmed. The space mission CoRoT \citep{esa3}
monitored the star in a serendipitous mode  since it is located in a
field contiguous to that of the open cluster NGC~2264, the main target of the
first short run in the anticenter direction (SRa01). 
We definitely
cross-identify \cep$\equiv$GSC~0746-01186$\equiv$ASAS~064135+0756.6$\equiv$2MASS
06413457 +0756396 when searching for Cepheids in the CoRoT data base 
(Poretti et al., in preparation).


\section{The analysis of \cep\, data}
\cep\, was observed continuously for 23.4~d 
from 2008 March 7 to March 31. 
The EXODAT catalog \citep{exodat} reports a spectral type of A5~III,
an E$_{B-V}$=1.1~mag (both from SED analysis),  and
a very low contamination level due to close stars, i.e., 0.004 in a range
from 0 to 1.
The ``CoRoT Variability Classifier" (CVC) automated supervised method \citep{cvc}
correctly suggests a classical or a double-mode Cepheid.
The measurements of \cep\, were performed in the 512-s mode from JD 2454533.4120 to
2454536.008 and then in the 32-s mode until the end of the observations, at JD 2454556.795.

\begin{figure}
\epsscale{1.00}
\plotone{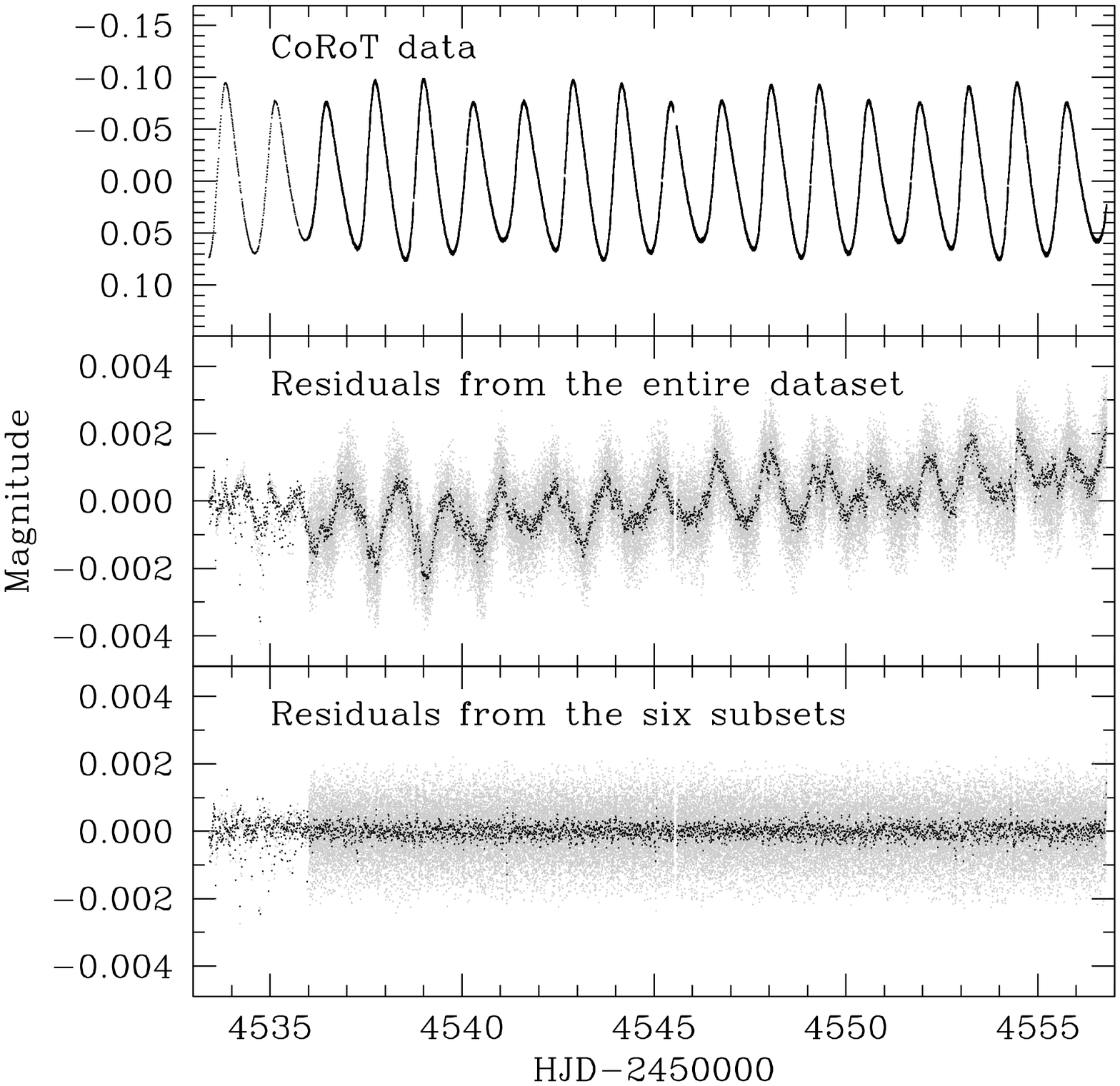}
\caption{{\it Top panel:} CoRoT photometry (white light) of \cep. {\it Middle panel:} residuals
from a solution calculated on the entire data set spanning 23.4~d.
{\it Bottom panel:} residuals from the solutions of six subsets spanning each 3.9~d.
Gray points: 32-s exposures; black points: 512-s exposures or 512-s averages.
}
\label{curves}
\end{figure}

All the CoRoT measurements were taken into account for the initial frequency analysis.
\ep{The outliers were removed by means of a cross-check between the flags provided by
the reduction pipeline and a visual inspection of the light curve. 
Outliers are for most measurements that suffer from hot pixels during the passage
through the South Atlantic Anomaly.}
The final time series is composed of 52,148 measurements:
they were obtained in the chromatic mode, but here we discuss the white
measurements only since we are interested in a detailed frequency analysis requiring a high
signal-to-noise ratio (S/N). 
The high-precision, high duty-cycle CoRoT photometry provides us with
an excellent representation of the beating between two commensurable periods 
(Fig.~\ref{curves}, top panel).
\ep{The light curve is a succession of two ``bright" maxima followed by two  ``faint" ones.}
The four maxima span the short beating period of 5.16~d=4$P_1$=5$P_2$.

\begin{figure}
\plotone{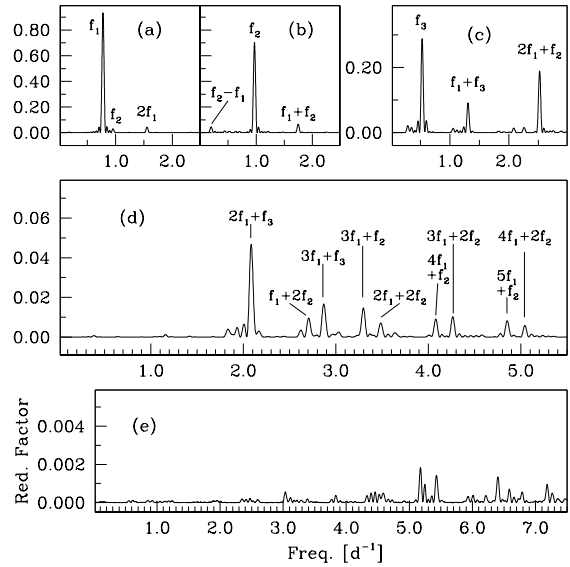}
\caption{\ep{Frequency analysis of the 
time series composed of the 512-s and 32-s measurements after subsequent prewhitenings.}
Panels {\it a} and {\it b}: detection
of \fu, \fdue, and combination terms. Panel {\it c}: detection of \ft\, and of the
combination term \fu+\ft. Panel {\it d}: detection of a subset of the combination
terms between \fu, \fdue, and \ft. Panel {\it e}: power spectra of the residuals
obtained from six subsets.
}
\label{spectra}
\end{figure}

The frequency analysis was performed by means of the iterative sine-wave, least-squares
fitting method \citep{vani}. It was applied to \ep{both}the original CoRoT time series composed of
32-s and 512-s exposures and to a new data set obtained by grouping 16 consecutive 32-s measurements
into a single 512-s point. The frequency values were refined by the MTRAP algorithm \citep{mtrap}, allowing
us to search for the best fit by keeping the values of the harmonics and the combination terms
locked to the independent frequencies. 
The analysis of the CoRoT data identified the two frequencies corresponding to the periods
already known \ep{(Fig.~\ref{spectra}, the power spectra from the original time series are shown)}. 
The peak at \fu=0.776~\cd\, is by far the highest
in the power spectrum, but 2\fu\, and \fdue=0.970~\cd\, are also immediately noticeable (panel {\it a}).
The \fdue\, peak and the combination terms \fu+\fdue\, and \fdue$-$\fu\, were detected 
after prewhitening the data with \fu\, and its harmonics (panel {\it b}). The four terms \fu, \fdue, \fu+\fdue,
and \fdue$-$\fu\ are the most common in the solutions of the light curves of double-mode Cepheids obtained
from ground-based data \citep{pardo}.
We expected that the subsequent analysis of the CoRoT time series 
would disclose the rich ensemble of combination terms between the harmonics of \fu\, and \fdue.
Therefore, it was a great surprise to detect a new independent term \ft=0.529~\cd, at the next step
(panel {\it c}). 
The presence of the \fu+\ft\, peak was particularly important since it immediately
ruled out that \ft\, was due to the variability of a close, unresolved star in the CoRoT mask or
to a binary companion. 
After this, 
the combination terms between the harmonics of \fu, \fdue, and \ft\, were detected step-by-step,
as shown in panel {\it d} for a subset of nine of them.
The final solution of the CoRoT light curve was provided by the independent terms \fu=0.776006 $\pm$ 0.000002~\cd, 
\fdue=0.969787 $\pm$ 0.000016~\cd, and \ft=0.529632 $\pm$ 0.000143~\cd,
their harmonics (up to 7\fu, 2\fdue, and 2\ft) and a set of 23 other combination terms, for a total
of 34~components.  
\ep{Table~\ref{sol} lists the ordered structure of the combination terms and
the least-squares solution obtained from the time series composed of the 512-s and 32-s
 measurements, which  leaves  a residual r.m.s. of 0.0010~mag.}
The error bars on the amplitudes $\sigma(A)=6\cdot10^{-6}$~mag and on the phases $\sigma(\phi_i)=\sigma(A)/A_i$
can be immediately derived following \citet{dsn}.

The discovery of \ft\, opened the possibility that other modes were excited and we performed an additional 
analysis to find them. 
The power spectra of the entire data set show some peaks close to \fu\ and, much less relevant,  to \fdue.
The highest one was at  $f$=0.728~\cd\, with an amplitude 0.008~times that of \fu.
The nature of these peaks is ambiguous. They
could have a stellar origin and be due to a long-term modulation of the main oscillations, observed in some
$1O/2O$ Cepheids \citep{pavel}. However, we cannot rule out the possibility that this long-term effect is actually due to satellite
drift and/or detector aging.
The analysis of the light curve of the residuals helped us to clarify this point. 
\ep{After the prewhitening with 34~components the data show residual oscillations, up to $\pm$2~mmag
(Fig.~\ref{curves}, middle panel).
The amplitudes of the peaks found in the frequency analysis of the residuals are smaller than 0.1~mmag. An
erratic nature could be  more plausible, but not fully convincing.}
Therefore, we calculated a new set of residuals by subdividing
the 23.4-d time baseline into six contiguous subsets spanning 3.9~d each. The frequency values
were kept fixed, but the amplitudes and phases were recalculated for each subset. Then,
the residuals obtained from these subsets were merged. The residual light curve thus obtained is almost
flat and the erratic oscillations have completely disappeared (Fig.~\ref{curves}, bottom panel).
We can infer that the erratic oscillations are  due to 
instrumental effects that are affecting the CoRoT photometry  in such a way
that the usual technique of prewhitening was not able to clean \citep[see also the case of
CoRoT 101155310; ][]{hads}.
\ep {The frequency analysis of the residuals after subtracting the 34 components from the
six subsets (Fig.~\ref{spectra},
panel {\it e}) did not detect any  peak below 1.0~\cd, where we expected the independent modes we were 
looking for. }

Therefore, we can conclude that after \ft\, no other independent mode is
detectable in the light curve of \cep.
\ep{The noise level of the power spectrum of the residuals (Fig.~\ref{spectra}, (panel {\it e})
is around 0.004~mmag in the 0-3~\cd\, region and
increases to 0.012~mmag in the 5.0-7.0~\cd\, region.
Some  of the peaks visible in
the 5.0-7.0~\cd\, region are close to combination terms (e.g., 6\fu+\ft=5.18~\cd, 7\fu=5.43~\cd,
7\fu+\fdue=6.40~\cd,  6\fu+2\fdue= 6.59~\cd), but their amplitudes
(0.028, 0.040, 0.028, and 0.030~mmag, respectively)
are below the heuristically accepted threshold of S/N=3.5 \citep{fgvir}.}

\ep{For the sake of completness, we successfully verified that the \ft\, component was also detectable in the
CoRoT data before removing the outliers and that
the solution of  the time series composed of the grouped 512-s points supplies the same combination terms
of the solution listed in Table~\ref{sol}.}

\section{Discussion}
The analysis of the data of \cep\, returned three independent periods:
$P_1$=1.2886~d, $P_2$=1.031140~d, and $P_3$=1.888538~d. This triplet is completely
new among galactic Cepheids. The period ratios are
$P_1/P_2$=0.8002, $P_2/P_3$=0.546, and $P_1/P_3$=0.682.
\subsection{The Petersen Diagram}
\begin{figure}
\plotone{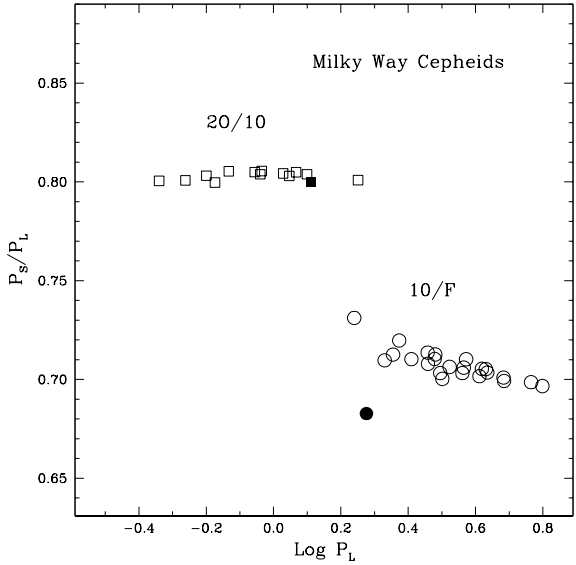}
\caption{Petersen diagram of galactic double-mode Cepheids
(open circles, $F$ and $1O$ stars; open squares, $1O$ and $2O$ pulsators).
The triple-mode Cepheid \cep\, is indicated with filled symbols.
}
\label{dm}
\end{figure}

Figure~\ref{dm} shows the
Petersen diagram of the  galactic double-mode Cepheids. There is  little doubt
on the fact that $P_1$ and $P_2$ can be typified with the $1O$ and $2O$ modes,
as is done in the galactic, LMC, and SMC double-mode Cepheids.
$P_3$ is giving us more trouble. The ratio 0.68 has been
observed in two double-mode LMC Cepheids \citep{soszinski2008a} and in six triple-mode
Cepheids ($1O/2O/3O$) in both the Magellanic Clouds \citep{pavel}.
The related  periods  were identified with those of the  $3O$ and $1O$ modes. However, this is not the case for
\cep: since $P_3$ is longer than $P_1$, it should be the $1O$ mode and $P_1$ the $3O$ one, 
contradicting the previous robust identification of $P_1$ as the $1O$ mode.

The fact that $P_3$ is much longer than $P_1$ and $P_2$ naturally suggests its identification with
the $F$ radial mode. However, the corresponding ratio $1O/F$=0.682 is
lower than the usual one \citep[0.694-0.746; ][]{pavel}. Moreover, Fig.~\ref{dm} shows how the
$1O/F$ ratio is expected to  increase toward short $F$-periods, thus  making  the 0.682 ratio
of \cep\, still more peculiar.
There is only one example of a ratio of 0.68 typified as a $1O/F$ one, i.e.,  J045917-691418
($P_0$=3.08~d and $P_1$=2.10~d) in
the LMC \citep[it is the apparently lowest outlier in Fig.~3 of ][]{eros}.
The unusual ratio is explained in terms of an high-metallicity of
$Z=0.030$ \citep[see also Fig.~3 in][]{buchlerszabo}, while the mean metallicity of LMC double-mode Cepheids is 0.004
\footnote{We cannot rule out
that J045917-691418 is actually a $3O/1O$ pulsator since the period ratio agrees very well with those of
the double-mode and triple-mode Cepheids recalled above. We verified that a $F$-period of
3.08/0.715=4.31~d matches the $P-L$ relation as well as 3.08~d.}.
The fundamental period of \cep\, is shorter than that of J045917-691418 and
a still higher metallicity is necessary to include the $(\log P,P_1/P_0)$ location of \cep\, between the limits
where $F$ and $1O$ are both unstable \citep{buchlerszabo}.

Therefore, if $P_3$ is actually the $F$ mode, \cep\, is a very particular star: not only
the unique $F/1O/2O$ Cepheid in the Galaxy but also one with an unusual $1O/F$ ratio.
In any case, \cep\, is the triple-mode Cepheid showing the longest periods, both in the Galaxy and
in the Magellanic Clouds.
In the context of the galactic Cepheids, it should be noted that $P$=1.8~d seems to be the shortest $F$ period among
the $1O/F$ double-mode stars and the longest $1O$ period among the $2O/1O$ ones (Fig.~\ref{dm}).

\subsection{The Analysis of the Fourier Parameters}
The Fourier decomposition could be used to disentangle the matter, since
the first harmonics were found for the three periods of \cep.
We calculated
the Fourier parameters $R_{ij}=A_i/A_j$
and $\phi_{ij}=j~\phi_{i}-i~\phi_{j}$
from the amplitude ($A_i$) and phase ($\phi_i$) coefficients of the $i$ and $j$
harmonics of the three independent frequencies \fu, \fdue, and \ft.

\ep {The parameters \fdu=4.1480$\pm$0.0006~rad and \rdu=0.2108 $\pm$ 0.0001 of \fu\, are
in excellent agreement with those of the $1O$ modes of the $1O/2O$ Cepheids.
In a similar way, the Fourier parameters of \fdue\, (\fdu=5.69$\pm$0.05~rad, \rdu=0.014$\pm$0.001)
are exactly located among those of the $2O$ modes 
\citep[see Figs.~2 and 3 in ][for the LMC Cepheids]{alcock}.
We conclude that \cep\, does not show any particularity as a $1O/2O$ double-mode pulsator.}


\ep{The situation is slightly different when we consider \ft.
The \fdu=4.05$\pm$0.12~rad value observed for \ft\,
is on the extension of the $F$-mode progression
\citep[see Fig.~4 in ][]{pardo}, but \rdu=0.047$\pm$0.005 is a
small value for the $F$-mode, usually greater than 0.20.}
 V371~Per is another double-mode Cepheid where
the \rdu=0.15 of the $F$-mode  is below the 0.20 limit and
smaller than the \rdu=0.22  of the $1O$-mode \citep{v371per}.
We can suppose that the small amplitude of the $F$-mode oscillation
reduces the \rdu\, value. It is also noteworthy that V371~Per shows an unusual high
$1O/F$ period-ratio (0.731), explained in terms of a metal deficiency ($-1<$[Fe/H]$<-0.7$).
Therefore, it seems that given stars could deviate from the most common ratios.

Another possibility is to identify \ft\, as a nonradial mode. Nonradial modes
have been detected in $1O$ LMC Cepheids and $F/1O$ double-mode LMC Cepheids, always closely
to the $1O$ mode \citep[$\Delta f<0.13$~\cd; ][]{pavel}. If we suppose a
canonical $F/1O$ frequency ratio of 0.72, we have $F=1O\cdot0.72$=0.559~\cd. The difference
$0.559-0.529=0.030$~\cd\, is still in the range where we can think of a resonance.
However, also under these assumptions \cep\, remains an unique and challenging
case since we have to model the excitation of a
nonradial mode close to the expected --~but not observed~-- $F$ radial mode.

\subsection{The Location of \cep\, in the Milky Way}
The $P-L$ relation for galactic Cepheids $M_V=-2.999(\pm0.097)\,\log P-0.995(\pm0.112)$ \citep{pl}
supplies similar $M_V$ values if we consider \ft=0.529~\cd\, or $0.776\cdot0.72=0.559$~\cd\, as
the frequency of the $F$ radial mode, i.e., $M_V=-1.82\pm0.12$ and $M_V=-1.75\pm0.12$, respectively.
The color excess $E_{B-V}$=1.1 \citep[EXODAT; ][]{exodat} yields $A_V=3.2\,E_{B-V}$=3.5,
\ep{$R$=3.2 from \citet{tammann}.}
The distance modulus $V-M_V=12.5+1.8-3.7=10.8$ and the galactic coordinates
($l$=204$\fdg$7407, $b$=+01$\fdg$4386)
put \cep\,  at 1.4\,kpc from the Sun, behind and farther than NGC~2264 (9.3; 202$\fdg$936 +02$\fdg$196).
The large color excess seems confirmed by available photometry: $V$=12.50 \citep{asas} and
$B$=13.6 (GSC~2.2 catalog). Smaller values
close to $E_{B-V}$=0.4 have been measured in that direction \citep{ebv}.
Such a value would push \cep\, at 4.0\,kpc from the Sun. 
\ep{ In any case, the
resulting  galactocentric distances   $R_G$=9.8~kpc ($E_{B-V}$=1.1)
or $R_G$=12.2~kpc ($E_{B-V}$=0.4)
or any intermediate one  obtained  from 0.4$<E_{B-V}<1.1$~mag)}
place \cep\, on
the extreme ``outer arm", not far from the B0\,II star HD~43818, i.e., in an environment known to
be metal rich \citep{outerarm}.


Therefore, the very high metallicity could cause the $1O/F$ period ratio to decrease in the opposite
sense of what the low metallicity probably does for V371~Per \citep{v371per}.
The fact that the $2O/1O$ ratio is normal is not
surprising, since the metallicity affects mainly the $1O/F$ ratio \citep{beaulieu, eros}.
Other evolutionary effects linked with the particular galactic location, as the first crossing
of the instability strip of a young object or an anomalous mass with respect to other double-mode
Cepheids, can contribute to the unusual period ratio.

\section{Conclusions}

The intensive CoRoT monitoring of \cep\, discovered a unique case among triple-mode
galactic Cepheids. The 0.682 ratio between the two longest periods is quite unusual.
If interpreted as  a $1O/F$ ratio, the fact that \cep\, belongs 
to  a metal-rich environment like the ``outer arm" of the Milky Way could explain the value.
We also note that the periods of \cep\, are much longer than those of the triple-mode 
Cepheids detected in the galactic bulge.
If not a radial mode, the excitation of an isolated nonradial mode 
with a period much longer than that of the $1O$ mode is also unusual for high-amplitude pulsators. 
\ep{As expected \citep[e.g., ][]{moskdziemb,pavel}, 
the discovery of  a new member of the rare class of triple-mode Cepheids set new  
observational constraints on the 
stellar parameters of these variables and, in a more wide context, on their evolutionary models.}

\acknowledgments
The CoRoT space mission has been developed and operated by CNES, with contributions
from Austria, Belgium, Brazil, ESA (RSSD and Science Program), Germany, and Spain.
This research has made use of the ExoDat Database, operated at LAM-OAMP,
 Marseille, France, on behalf of the CoRoT/Exoplanet program.
The present study has used the SIMBAD data base operated at the Centre
de Donn\'ees Astronomiques (Strasbourg, France).

\clearpage



\begin{table}
\begin{center}
\caption{
\ep{The identification of the 34 significant frequencies detected in the data of \cep.
The amplitudes and phases obtained from
the original 32-s and 512-s measurements are listed\label{sol}.}}
\smallskip
\smallskip
\begin {tabular}{rrrr}
\tableline\tableline
\noalign{\smallskip}
\multicolumn{1}{c}{ID} & \multicolumn{1}{c}{Frequency} & \multicolumn{1}{c}{Amplitude} & \multicolumn{1}{c}{Phase} \\
& \multicolumn{1}{c}{$\mathrm{[d^{-1}]}$} &\multicolumn{1}{c}{[mag]} &\multicolumn{1}{c}{[0,2$\pi$]}  \\
\noalign{\smallskip}
\hline
\noalign{\smallskip}
\fu         &     0.776006 &   0.071792 &  0.6504 \\
\fdue       &     0.969787 &   0.009816 &  3.8558 \\
\ft         &     0.529632 &   0.001181 &  1.5836 \\
2\fu        &     1.552012 &   0.015132 &  5.4488 \\
3\fu        &     2.328018 &   0.003827 &  3.8502 \\
4\fu        &     3.104024 &   0.000723 &  2.4672 \\
5\fu        &     3.880030 &   0.000198 &  0.1619 \\
6\fu        &     4.656036 &   0.000089 &  4.7833 \\
2\fdue      &     1.939574 &   0.000137 &  0.8405 \\
2\ft        &     1.059264 &   0.000056 &  0.9304 \\
\fdue--\fu   &     0.193781 &   0.002828 &  1.0666 \\
\fu+\fdue   &     1.745793 &   0.003041 &  2.9196 \\
2\fu+\fdue  &     2.521799 &   0.001057 &  1.7247 \\
3\fu+\fdue  &     3.297805 &   0.000160 &  2.1453 \\
4\fu+\fdue  &     4.073811 &   0.000116 &  1.6367 \\
5\fu+\fdue  &     4.849817 &   0.000101 &  0.3820 \\
6\fu+\fdue  &     5.625823 &   0.000066 &  5.8167 \\
2\fu--\fdue  &     0.582225 &   0.000324 &  5.1539 \\
3\fu--\fdue  &     1.358231 &   0.000066 &  0.6502 \\
4\fu--\fdue  &     2.134237 &   0.000056 &  3.2205 \\
\fu+2\fdue  &     2.715580 &   0.000162 &  6.0254 \\
2\fu+2\fdue &     3.491586 &   0.000104 &  5.5258 \\
3\fu+2\fdue &     4.267592 &   0.000104 &  4.8089 \\
4\fu+2\fdue &     5.043598 &   0.000080 &  3.8391 \\
5\fu+2\fdue &     5.819604 &   0.000055 &  2.5362 \\
2\fdue--2\fu &     0.387562 &   0.000056 &  1.5524 \\
4\fu+\ft    &     3.633656 &   0.000061 &  0.9406 \\
\fu+\ft     &     1.305638 &   0.000699 &  6.0973 \\
2\fu+\ft    &     2.081644 &   0.000286 &  4.4819 \\
3\fu+\ft    &     2.857650 &   0.000153 &  2.8912 \\
\fu--\ft     &     0.246374 &   0.000271 &  4.3864 \\
\fdue+\ft   &     1.499419 &   0.000192 &  2.9091 \\
\fdue--\ft   &     0.440155 &   0.000051 &  6.0342 \\
\fu+\fdue+\ft&    2.275425 &   0.000069 &  1.9800 \\
\noalign{\smallskip}
\multicolumn{4}{l}{N measurements: 52148}\\
\multicolumn{4}{l}{$T_0$= HJD 2454546.3070}\\
\multicolumn{4}{l}{Residual r.m.s.~~~ 0.00104 mag}\\
\tableline
\end{tabular}
\end{center}
\end{table}
\end{document}